\documentclass[10pt,prl,aps,showpacs,twocolumn]{revtex4}
\usepackage[caption=false]{subfig}
\usepackage{amssymb}
\usepackage{amsmath}
\usepackage{commath}
\usepackage{graphicx,bm}
\usepackage{verbatim}
\usepackage[usenames]{xcolor}
\usepackage{pdfpages}
\usepackage{hyperref}
\usepackage[normalem]{ulem}
\usepackage{blindtext}

\def\be#1\ee{\begin{equation}#1\end{equation}}
\def\ba#1\ea{\begin{align}#1\end{align}}

\newcommand{\eqqref}[1]{Eq.\,\eqref{#1}}	
\newcommand{\figgref}[1]{Fig.\,\ref{#1}}

\newcommand{\bk}	{{\bm k}}

\newcommand{\bq}	{{\bm q}}

\newcommand{\bl}	{{\bm \lambda}}
\newcommand{\bx}    {{\bm x}}
\newcommand{\by}    {{\bm y}}

\newcommand{\s}	    {\sigma}
\newcommand{\w}	    {\omega}

\newcommand{\veps}  {\varepsilon}

\begin{document}
\title{Many-body fermionic excitations in Weyl semimetals due to elastic gauge fields}
\author{E.C.I. van der Wurff}
\email{e.c.i.vanderwurff@uu.nl}
\affiliation{Institute for Theoretical Physics, Princetonplein 5, 3584 CC Utrecht, the Netherlands}
\author{Alberto Cortijo}
\email{alberto.cortijo@csic.es}
\affiliation{Materials Science Factory, Instituto de Ciencia de Materiales de Madrid, CSIC, Cantoblanco, 28049 Madrid, Spain.}

\date{\today}
\begin{abstract}
We study the single-particle spectrum of three-dimensional Weyl semimetals taking into account electron-phonon interactions that are the result of straining the material. We find that a well-defined fermionic excitation appears in addition to the standard peak corresponding to quasiparticle states as suggested by Landau-Fermi liquid theory. Contrary to the case of Dirac systems interacting via the Coulomb interaction, these satellite peaks appear even at lowest order in perturbation theory. The new excitations are anisotropic, as opposed to the single-particle spectrum, and their behavior is dictated by the Debye frequency, which naturally regulates the electron-phonon coupling.
\end{abstract}

\maketitle
\textit{Introduction.---} Quite generically, equilibrium and transport properties of many-body metallic systems are accounted for by the Landau theory of the Fermi liquid. This theory is based on the notion of \emph{quasiparticles}: single-particle fermionic excitations that are in one-to-one correspondence to the non-interacting fermionic states. In this regard, the single-particle spectral function $\rho(\omega,\bm{k})$ provides information about how good the quasiparticle approximation is in an interacting system. For non-interacting fermions the spectral function is given by a Dirac delta function, $\rho^0(\omega,\bm{k})=\delta(\hbar\omega+\mu-\varepsilon_{\bm{k}})$, where $\varepsilon_{\bm{k}}$ is the dispersion relation of the particle and $\mu$ the chemical potential. Upon including interactions, Landau's Fermi liquid theory states that, because the damping rate of the quasiparticles scales quadratically with the energy at the Fermi level, the spectral function $\rho^0(\omega,\bm{k})$ becomes a Dirac delta function that is weighted by a parameter $0<Z<1$.

It was realized a long time ago \cite{Lundqvist1967,Hedin67} that the actual spectral function of an interacting fermionic system is richer than the one suggested by Landau-Fermi liquid theory. Apart from the existence of a well-defined quasiparticle peak, extra weakly-damped, i.e. long-lived, fermionic satellite excitations appear due to the coupling of bare fermions with propagating bosonic degrees of freedom. Such satellite excitations, currently called \emph{plasminos}, were theoretically reported in the context of the quark-gluon plasma \cite{Klimov82,Weldon89}, and more generically in hot or dense relativistic plasmas \cite{Baym92,Ollitrault1993,Manuel2000}. In condensed matter physics, they are referred to as \emph{plasmarons}, and they have been theoretically studied in graphene \cite{MacDonald2008,Sensarma2011,Carbotte2011,Carbotte2012} and Bismuth \cite{Giamarchi2011}. Experimental evidence of these emergent modes has been found using angle-resolved photoemission spectroscopy (ARPES) and optical measurements in graphene \cite{MacDonald2010} and Bismuth \cite{VanderMarel2007}. These satellite fermionic excitations are not just of purely theoretical interest. They modify the transport and optical properties of the electronic system \cite{Giamarchi2011,Sadooghi2014}: in the case of plasminos, it is well known that they have an opposite helicity-to-chirality ratio as compared to the quasiparticle excitations \cite{Weldon82}.

In the context of condensed-matter physics, Weyl semimetals are materials that host low-energy quasiparticles that linearly disperse \cite{VishwanathRMP}. These quasiparticles also have a chirality and have similar kinematics to chiral ultra-relativistic particles. It is therefore interesting to analyze the many-body fermionic spectrum in these systems and see how many-body quantum effects interfere with ultra-relativistic (or pseudo-relativistic) invariance in solid-state systems. 

Despite their appealing similarities to ultra-relativistic systems, we have to keep in mind that Weyl semimetals are condensed matter systems in which the transverse electromagnetic degrees of freedom are not important compared to the collective plasmon mode \cite{Neto2012}. Moreover, in three-dimensional systems, longitudinal and transverse plasmonic modes become gapped, with the gap set by the plasma frequency. Effects associated to electron-plasmon interactions are therefore weaker than in the two-dimensional counterpart \cite{dassarma2015a,dassarma2015b,Fu2018}. Phonons, on the contrary, remain gapless even after including many-body effects in three dimensions. Besides, electron-phonon interactions appear in Weyl semimetals in the form of elastic gauge fields \cite{Vozmediano2015,Grushin2019}, so they become a promising alternative to study the physics of emergent fermionic degrees of freedom different from the quasiparticles, even at lowest order in perturbation theory \cite{Klimov82,Baym92,Ollitrault1993,Ollitrault1997,Manuel1997,Manuel2000}.

In the present work we show that novel fermionic excitations appear in Weyl semimetals due to electron-phonon couplings that can be expressed in terms of elastic gauge (vector) fields. A distinctive feature of the electron-phonon coupling is that extra energy/momentum scales appear in the problem. On the one hand there are the longitudinal/transverse phonon velocities, which are typically much smaller than the Fermi velocity of the electrons, and on the other hand there is the Debye frequency, which naturally acts as a cutoff for the electron-phonon interaction. Another interesting feature is that, as we will see, despite the isotropic nature of the original non-interacting electronic spectrum, the emergent fermionic modes will be anisotropic due the particular nature of the elastic vector fields in Weyl metals. Both the anisotropy and the presence of additional energy and momentum scales make the problem quite different from the situation in relativistic plasmas. 

\textit{Strain in Weyl metals.---}
We start by considering the minimal model for a time-reversal symmetry breaking Weyl metal. Setting $\hbar \equiv 1$, the Hamiltonian that describes two Weyl cones separated by $2\lambda_3\hat{\bm{z}}$ reads
\be
\mathcal{H}^{W}_{0,\chi}(\bm{k})=v\bm{\sigma}_{\perp}\cdot\bm{k}_{\perp}- \chi v_3(k_3-\chi\lambda_3)\sigma_3,\label{eq:Ham1}
\ee
with ${\bm k} = ({\bm k}_{\perp},k_3)$, ${\bm \sigma} = ({\bm \sigma}_{\perp},\s_3)$ Pauli matrices, $\chi=\pm$ the chirality of the cone and $v$ and $v_3$ Fermi velocities.

It was shown in Ref.\,\cite{Vozmediano2015} that when strain is considered, the lowest-order modification of the Hamiltonian in \eqqref{eq:Ham1} due to strain is $\delta\mathcal{H}[u]=v \chi\bm{\sigma}_{\perp}\cdot\bm{A}^{\text{el}}_{\perp} + v_3A^{\text{el}}_3\sigma_3$.
From this expression, we see that the presence of the chirality $\chi$ implies that the coupling between electrons and the strain tensor is through a chiral vector potential $\bm{A}^{\text{el}} = ({\bm A}^{\text{el}}_{\perp}, A_3^{\text{el}})$. That is, it couples with a different sign to the two Weyl nodes with opposite chirality. This so-called elastic gauge field can be expressed in terms of the microscopic tight-binding parameters leading to the low-energy Hamiltonian in \eqqref{eq:Ham1} and the linearized strain tensor $u_{ij} \equiv (\partial_iu_j  + \partial_j u_i)/2$ that itself is defined in terms of the displacement vector $u_i$, which measures the deviation from the equilibrium situation \cite{Vozmediano2015,LandauElasticity}. 

The influence of strain can also be derived from a more generic symmetry-based approach. In this line of approach, all possible scalars that can be constructed from the available scalars, vectors and tensors are added to the Hamiltonian for the unstrained material. At first order in the strain tensor, the strain-induced Hamiltonian $\delta\mathcal{H}[u]$ for one cone contains three possible terms \cite{Vozmediano2018}, i.e.,
\be
\delta\mathcal{H}[u] = v g_0 \sigma_0 \text{Tr}[u] + v g_1 \sigma_i u_{ij}\lambda_j + v g_2 \sigma_i\delta_{ij}\lambda_j\text{Tr}[u], \label{eq:HamSym}
\ee
with $g_{0,1,2}$ coupling constants. The first term, proportional to the identity matrix $\sigma_0$, is the standard electron-phonon coupling coming from the changes induced in the electronic density due to changes in the volume, $\delta V/V =  \text{Tr}[u]$. The Hamiltonian given by \eqqref{eq:HamSym} gives rise to different types of electron-phonon coupling. These couplings can be derived by writing the strain tensor $u_{ij}$ in terms of the displacement $u_i$ and subsequently quantizing the latter to yield phonons. Phonons come in two types: longitudinal and transverse. By construction, only longitudinal phonons contribute to the trace in the first and third term of \eqqref{eq:HamSym}, whereas the second term receives contributions from both longitudinal and transverse phonons. 

The coupling between transverse phononic degrees of freedom and electrons is similar to what happens in QED (quantum electrodynamics), where electrons couple to transverse gauge degrees of freedom. However, there is a crucial difference: in ultrarelativistic QED both the gauge and fermionic degrees of freedom have the same velocity, and the transverse modes are gapped as the longitudinal collective plasmonic mode \cite{Bellac2000}. In the case of phonons, in contrast, the phonon velocity is much smaller than the Fermi velocity, thereby resembling the case of sound modes in plasmas due to ionic collective modes. These modes do not develop a gap but they do attain a strong renormalization of the sound velocity. These observations place the vector electron-phonon coupling in Eq.(\ref{eq:HamStart}) in an unique situation not addressed before in condensed matter, nor in relativistic plasmas.

We therefore proceed by focussing on the second term in \eqqref{eq:HamSym}. The corresponding second-quantized continuum Hamiltonian $\delta \mathcal{H}_2[u]$ reads in momentum space
\be
\!\delta \mathcal{H}_2[u] =\frac{iv_Fg_1}{2}\!\int_{\bk,\bq}\! \psi^{\dagger}_{\bm{k}+\bm{q}}\sigma_i\lambda_{j}(q_{i}\delta_{jr}+q_{j}\delta_{ir})u_{r}(\bm{q})\psi_{\bk},\! \label{eq:HamStart}
\ee
where $\int_{\bk} \equiv \int \text{d}^3\bk/(2\pi)^3$,  $\psi_{\bk}^{\dagger}$ ($\psi_{\bk}$) is the fermionic creation (annihilation) operator and we denoted the Fermi velocity by $v_F$. We wrote \eqqref{eq:HamStart} explicitly in terms of the displacement field $u_r(\bq)$ using the definition of the strain tensor. Quantizing the displacement field, we find the electron-phonon vertex $g^{\alpha}_{\bm{q}}$, i.e.,
\be
g^{\alpha}_{\bm{q}} = \frac{i}{2}\frac{v_F g_1}{\sqrt{2\rho \omega_{\alpha}(\bm{q})}}\sigma_{i}\lambda_{a}(q_i\delta_{ar}+q_a\delta_{ir})\hat{e}^{\alpha}_{r}, \label{eq:Vertex}
\ee
with $\rho$ the mass density of the material under consideration, $\w_{\alpha}(\bq) = c_{\alpha}|\bq|$ the dispersion relation of the longitudinal ($\alpha =L$) and transverse acoustic phonons ($\alpha =T$) and $c_{\alpha}$ the corresponding speeds of sound. Furthermore, the polarization vectors $\hat{e}^{\alpha}_r$ form a basis in momentum space and obey the identity $\sum_{\alpha}\hat{e}^{\alpha}_i\hat{e}^{\alpha}_j=\delta_{ij}$, with $\hat{e}^L_i = \hat{q}^i$.

It is important to stress that the electron-phonon coupling in \eqqref{eq:Vertex} follows from very generic elasticity theory. It therefore holds in any Weyl semimetal. Contributions from optical phonons, in contrast, are rather material and model-dependent. Additionally, due to their defining energy-gap at the $\Gamma$-point, we expect the contributions from optical phonons to be negligible. We therefore do not consider the effect of optical phonons here and proceed with the computation of the self-energy due to the coupling of the electrons to the acoustic phonons. 

\textit{Electronic self-energy.---} 
For simplicity we consider an isotropic Weyl cone with dispersion relation $\veps_{\bk} = v_F|\bk|$. Furthermore, we treat the acoustic phonons within the Debye model such that upon crossing the Debye frequency $\w_D$, the phonon density of states becomes zero. The Matsubara Green's function for the acoustic phonons therefore reads \cite{Abrikosov2012}
\be
D_0^{\alpha\alpha'}(i\omega_m,\bm{q})=\frac{2\w_{\alpha}(\bq)}{(i\w_m)^2 - \w_{\alpha}^2(\bq)}\vartheta(k^{\alpha}_D - |\bq|)\delta_{\alpha\alpha'}\label{eq:GreenPhonon},
\ee
with $i\w_m$ a bosonic Matsubara frequency and $k^{\alpha}_D = \w_D/c_{\alpha}$ the Debye wavenumber. Upon integrating out the phonons, we obtain the one-loop electron self-energy \cite{Sadovskii2006}. At $T=0$ the retarded self-energy reads \cite{Ollitrault1993}
\ba
\!\!\!\!\Sigma(\omega,\bm{k})&=\!\sum_{\substack{\alpha \in \{L,T\} \\u = \pm}}\int_{k^{\alpha}_D > |\bq|} g^{\alpha}_{\bm{-q}}\bigg[\frac{\Lambda_{-u}(\bm{k}-\bm{q})}{\omega^+ + u\veps_{\bm{k}-\bm{q}}+u\omega_{\alpha}(\bq)} \nonumber \\
&\phantom{=}\qquad\qquad+ \frac{u\Lambda_{+}(\bm{k}-\bm{q})\vartheta(\mu - \varepsilon_{\bm{k}-\bm{q}})}{\omega^+-\varepsilon_{\bm{k}-\bm{q}}+u\omega_{\alpha}(\bq)}\bigg]g^{\alpha}_{\bm{q}}, \label{eq:Selfenergy2}
\ea
with $\Lambda_{\pm}(\bk) \equiv (\s^0 \pm  {\bm\s}\cdot\hat{\bk})/2$ and $\w^+ = \w + i0$. Note that we absorbed the chemical potential in the frequency $\w$ and that the integral is rendered finite by the high-momentum cut-off provided for by the Debye wavenumber. The sum over the polarizations in \eqqref{eq:Selfenergy2} can be performed separately for the longitudinal and transverse phonons. The full self-energy is then given by the sum of the longitudinal and transverse contribution, i.e., $\Sigma(\w,\bk) \equiv \Sigma^L(\w,\bk)  + \Sigma^T(\w,\bk)$. For simplicity we will take $c_L = c_T\equiv c$ in the following. 

Before we compute the self-energy, it is instructive to introduce dimensionless variables by scaling all parameters on $|\bl|$, i.e., $\tilde{\w} \equiv \w/v_F |\bl|$, $\tilde{k} \equiv|\bk|/|\bl|$, $\tilde{\mu} \equiv \mu/v_F |\bl|$ and $\tilde{k}_D \equiv k_D/|\bl|$. The ratio of the dimensionless chemical potential and Debye wavenumber is given by $\tilde{\mu}/\tilde{k}_D = c \mu/v_F \w_D \equiv \veps \mu/\w_D$ in terms of the typically small parameter $\veps \equiv c/ v_F$. This is rather different from the situation in relativistic plasmas, where the constituent particles are photons and electrons, both moving at the speed of light. Another important difference between the case at hand and relativistic plasmas is that there is no natural Debye frequency scale for photons like we have here. It is therefore interesting to investigate the different possible regimes set by $\tilde{\mu} / \tilde{k}_D$. 

From now on we use these dimensionless variables and omit tildes. The dimensionless self-energy can be decomposed as
\be
\frac{\Sigma(\w,\bk)}{v_F |\bl|} = G_1\big[\Sigma_0(\w,\bk)\s_0 + \Sigma_{1}(\w,\bk)\bm{\s}\cdot\hat{\bk}\big], \label{eq:Selfenergy3}
\ee
with $G_1 \equiv g_1^2|\bl|^4 / 4 \rho v_F \veps$ a new dimensionless coupling constant. Interestingly, the original coupling constant $g_1$ is thus increased by a factor of $1/\veps \gg 1$.

After the scaling on $|\bl|$, the self-energy still depends on the direction $\hat{\lambda}^i$, as can be seen from the vertex in \eqqref{eq:Vertex}. To illustrate the resulting anisotropy in the self-energy, we consider the contribution of the longitudinal phonons to $\Sigma_1(\w,\bk)$ in some more detail. From \eqqref{eq:Selfenergy2} it can be expressed as
\ba
&\Sigma^L_{1}(\w,\bk) = \hat{\lambda}^i\hat{\lambda}^j\int_{k_D > |\bq|} \frac{\hat{q}^i\hat{q}^j}{|\bk-\bq|}H_-(\w,\bk-\bq,\bq) \nonumber\\
&\phantom{=}\qquad\qquad\qquad\times\big[(2\bk\cdot\bq - |\bq|^2)(\hat{\bk}\cdot\hat{\bq}) - |\bk||\bq|\big], \label{eq:Selfenergy4}
\ea
with $H_{-}(\w,\bk-\bq,\bq)$ a scalar function that will be specified later. The tensorial structure in \eqqref{eq:Selfenergy4} can be simplified by noting that the integrand is symmetric under $i\leftrightarrow j$. The integral thus has to yield a linear combination of $\delta^{ij}$ and $\hat{k}^i\hat{k}^j$. The latter leads to a contribution proportional to $(\hat{\bl}\cdot\hat{{\bm k}})^2$, whereas the former does not depend on the angle between $\bl$ and $\bk$. Clearly the resulting expression depends on the angle between $\bl$ and $\bk$ in a non-trivial way. 
\begin{figure*}[t!]
\includegraphics[scale=.63]{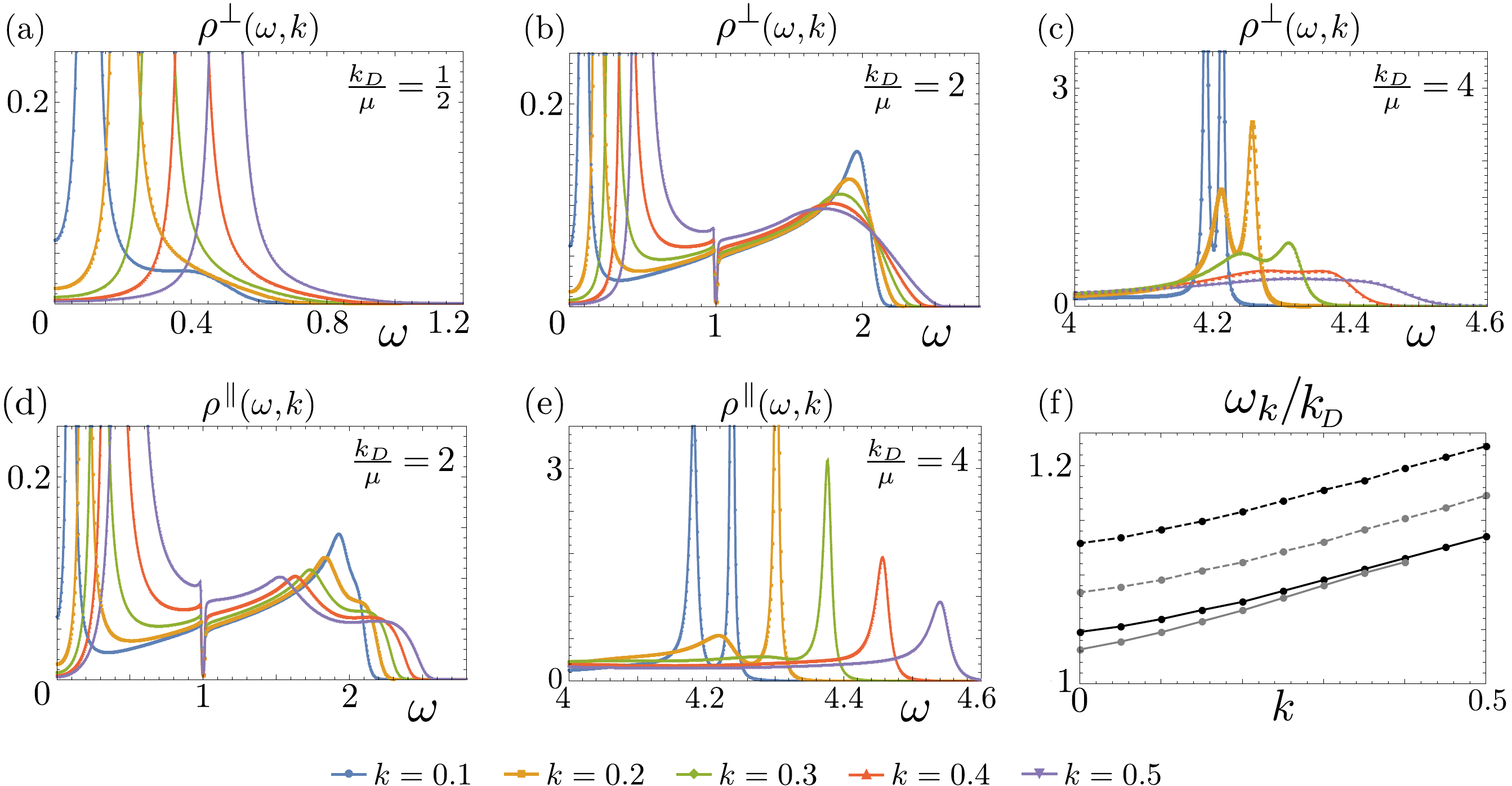}
\caption{(a)-(e) Plots of the spectral function $\rho(\w,\bk)$, as a function of $\w\equiv \w / v_F |\bl|$ and for five different values of $k \equiv |\bk|/|\bl|$. Figs.\,(a)-(c) are for the case $\bl\perp\bk$ and (d), (e) for $\bl\parallel\bk$. In (c) and (e) we zoom in on frequencies around the Debye frequency, but the quasiparticle peaks for smaller frequencies remain present. For all spectral functions we used $\veps = 1/100$ and $G_1 = 1.0$. Note that all spectral functions go to zero at $\w \equiv \w + \mu = 1$ (recall that we absorbed the chemical potential in the frequency). This is generically the case in an interacting electron liquid because the imaginary part of the self-energy goes to zero for $\w\to \mu$ \cite{Luttinger1961}. (f) Dispersion relation following from the location of the emergent peaks in the spectral functions from (e). Used parameters in this plot are (from bottom to top): $k_D/\mu = 3.5$ and $\veps = 1/100$ (gray), $k_D/\mu = 4.0$ and $\veps = 1/100$ (black), $k_D/\mu = 4.0$ and $\veps = 1/20$ (gray dashed), $k_D/\mu = 4.0$ and $\veps = 1/10$ (black dashed).}
\label{fig:Spectral}
\end{figure*}

In what follows we investigate the anisotropy of the self-energy in two distinct cases: $\bl\perp\bk$ and $\bl\parallel\bk$. The corresponding self-energy contributions are $\Sigma^{\theta}_{n}(\w,\bk)$, with $\theta = \perp,\parallel$ and $n=0,1$. They can be expressed as
\be
\Sigma_n^{\theta}(\w,\bk) = \frac{1}{8}\int_{k_D>|\bk+\bq|}\frac{f_n^{\theta}(\bk,\bq)H_{\eta_n}(\w,\bq,\bk+\bq)}{|\bk+\bq|^{2n+1}}, \label{eq:Selfenergy5}
\ee
with $\eta_0 = +, \eta_1 = -$. The function $H_{\pm}(\w,\bx,\by)$ reads
\ba
&\hspace{-.2cm}H_{\pm}(\w,\bx,\by) \equiv \frac{\vartheta(\mu - |\bx|)}{\w^+ - |\bx| + \veps|\by|} - \frac{\vartheta(\mu - |\bx|)}{\w^+ - |\bx| - \veps|\by|} \nonumber\\
&\phantom{=}\qquad\qquad+ \frac{1}{\w^+-|\bx|-\veps|\by|} \pm \frac{1}{\w^+ + |\bx| + \veps|\by|}. \label{eq:HCollective}
\ea
Finally, the functions $f_n^{\theta}(\bk,\bq)$ in \eqqref{eq:Selfenergy5} capture the angle-dependence and are functions of $|\bk|$, $|\bq|$ and $\bk\cdot\bq$. We give explicit derivations and expressions for them in the Supplemental Material.

\textit{Results.---} Using the expressions following from \eqqref{eq:Selfenergy5}, we calculate the imaginary part of the self-energy analytically. Subsequently, we obtain the real part by numerically integrating the corresponding Kramers-Kronig relation. To investigate the many-body behavior resulting from the self-energy, we calculate the dimensionless spectral function $\rho^{\theta}(\w,\bk) \equiv \rho^{\theta}_+(\w,\bk) + \rho^{\theta}_-(\w,\bk)$, with
\be
\!\!\rho^{\theta}_{\pm}(\w,\bk) =\frac{-G_1\text{Im}[\Sigma^{\theta}_{\pm}]/\pi}{\big(\w \mp k - G_1\text{Re}[\Sigma^{\theta}_{\pm}]\big)^2 + \big(G_1\text{Im}[\Sigma^{\theta}_{\pm}]\big)^2}, \label{eq:spectralfunc}
\ee
where $\Sigma^{\theta}_{\pm}(\w,\bk) \equiv \Sigma^{\theta}_0(\w,\bk) \pm \Sigma^{\theta}_1(\w,\bk)$. The spectral function obeys the sum rule $\int_{-\infty}^{\infty}\text{d}\w\,\rho^{\theta}(\w,\bk) = 2$, as a Weyl fermion has two degrees of freedom. We checked that the sum rule is satisfied for all numerically obtained spectral functions within $1\%$ accuracy. We plot the spectral function in \figgref{fig:Spectral} for several interesting cases and proceed to discuss the results presented there in detail.

We start with the case $\bl \perp \bk$. In \figgref{fig:Spectral}(a) we show $\rho^{\perp}(\w,\bk)$ for $k_D/\mu = 1/2$ and five different values of $|\bk| \equiv |\bk|/|\bl|$, taking small enough values such that the linear approximation around the Weyl nodes remains justified. As expected, we observe high quasiparticle peaks centered around $\w_{\bk} = \tilde{v}_F|\bk|$, with $\tilde{v}_F$ a renormalized Fermi velocity. There are however, no other satellite peaks present in this case. The situation changes when we increase the ratio $k_D/\mu$. In \figgref{fig:Spectral}(b) we show the case $k_D / \mu = 2$, for which a broad peak appears around the Debye frequency. As $k_D/\mu$ is increased even further, c.f.\,\figgref{fig:Spectral}(c), the peak centers around the new value for $k_D/\mu$ and becomes narrower and higher, signaling the emergence of a well-defined excitation. Interestingly, the peak is split in two for small $|\bk|/|\bl|$ and is damped out upon increasing $|\bk|/|\bl|$. 

The situation is more pronounced when $\bl\parallel \bk$. In this case, the behavior in the regime $k_D/\mu < 1$ is very similar to the behavior presented for the perpendicular case in \figgref{fig:Spectral}(a), so we refrain from showing a plot of it here. Upon increasing $k_D/\mu$, a distinct behavior from the case $\bl\perp\bk$ becomes clear, as can be seen in Figs.\,\ref{fig:Spectral}(d) and (e). Especially in \figgref{fig:Spectral}(e) it is clear that the peaks are even narrower and higher than in \figgref{fig:Spectral}(c). Again the peaks are centered around the Debye frequency $k_D$ and they disperse as $|\bk|/|\bl|$ is increased. In contrast to \figgref{fig:Spectral}(c) the damping of the peaks as $|\bk|/|\bl|$ is increased is much smaller, leading to a well-defined satellite peak at higher wavenumbers. The height of the satellite peak is for a large part determined by the ratio between the phonon and electron velocity $\veps \equiv c / v_F$. As $\veps$ is increased from, e.g., $1/100$ to $1/10$, the peaks become lower and broader, in addition to being located at higher frequencies. The corresponding plots are shown in the Supplemental Material.

In order to investigate the dispersive behavior, we collect the locations of the peaks for several values of $|\bk|/|\bl|$ and plot the corresponding dispersion relation $\w_{\bk}$, normalized on the Debye wavenumber $k_D$, in \figgref{fig:Spectral}(f). We plot the dispersion relation for different values of the ratio between the phonon and electron velocity and the Debye wavenumber $k_D$. Clearly the dispersion is linear for larger wavenumbers, but it seems to slightly change slope as the wavenumber goes to zero. From \figgref{fig:Spectral}(f) it is clear that the slope of the linear part of the dispersion is set by the phonon velocity via the dimensionless parameter $\veps$. Upon closer inspection of the last two terms in \eqqref{eq:HCollective}, it appears that the slope is set by $v_F + c = v_F(1 + \veps)$. Scaling the curves in \figgref{fig:Spectral}(f) on $(1+\veps)$ does not lead to a perfect collapse due to numerical inaccuracies.

\textit{Conclusion.---}
In summary, we have shown that gauge field-like electron-phonon coupling leads to novel satellite excitations in Weyl metals. The excitations have a rich phenomenology, depending on three parameters: i) the ratio between the Debye wavenumber and the chemical potential $k_D/\mu$, ii) the ratio between the phonon and fermion velocity $c/v_F$ and iii) the angle between the external wavenumber and the separation between the Weyl nodes, $\hat{\bl}\cdot\hat{\bk}$. It is also important to note that the transverse phonons contributed significantly to all the spectral functions in \figgref{fig:Spectral} and excluding them leads to very different results indeed.

Regarding the potential impact of higher orders in perturbation theory, we do not expect vertex corrections to significantly change the results presented here, as Migdal's theorem for massless fermions dictates that vertex corrections are surpressed by a factor of $c/v_F \ll 1$ \cite{DasSarma2014}. Finally, we note that the emergent fermionic modes found in this work are well defined (the imaginary part of the self-energy becomes much smaller than the real part at small momenta) when $c/v_F \ll 1$, so in principle it is possible to implement schemes based on kinetic theory to compute the impact of these excitations on transport and optical properties \cite{Blaizot93,Blaizot93a,Sadooghi2014}. A combination of optical, transport, and ARPES measurements could therefore be employed to show the existence of the new fermionic excitation experimentally.

\begin{acknowledgments}
It is a pleasure to thank M.\,A.\,H. Vozmediano, K. Landsteiner and H.\,T.\,C. Stoof for useful discussions. This work is supported by the Stichting voor Fundamenteel Onderzoek der Materie (FOM) and is part of the D-ITP consortium, a program of the Netherlands Organisation for Scientific Research (NWO) that is funded by the Dutch Ministry of Education, Culture and Science (OCW). A.C. acknowledges financial support through MINECO/AEI/FEDER, UE Grant No. FIS2015-73454-JIN and European Union structural funds and the Comunidad Aut\'onoma de Madrid 
(CAM) NMAT2D-CM Program (S2018-NMT-4511)
\end{acknowledgments}
\bibliography{phononpolariton}

\appendix
\onecolumngrid
\newpage
\section{Supplemental Material: derivation of the self-energy and additional plots}
\noindent In this Supplemental Material we first derive \eqqref{eq:Selfenergy2}. Subsequently we derive the expression in \eqqref{eq:Selfenergy5} and the corresponding functions $f_n^{\theta}(\bk,\bq)$. Finally, we show some additional plots of the spectral function to support the discussion of \figgref{fig:Spectral} in the main text.
\subsection{Matsubara sum}
\noindent Upon integrating out the phonons, the full expression for the self-energy is given by
\be
\Sigma(i\omega_m,\bm{k})=-\frac{1}{\beta}\sum_{i\w_n,\alpha,\alpha'}\int_{|\bq|<k_D^{\alpha}}g^{\alpha}_{\bm{-q}}G_{0}(i\w_{m}-i\w_{n},\bm{k}-\bm{q})g^{\alpha'}_{\bm{q}}D_0^{\alpha\alpha'}(i\w_{n},\bm{q}). \label{app:self}
\ee
The Matsubara Green's function for a Weyl fermion with an isotropic Weyl cone and positive chirality can be written as
\be
G_{0}(i\omega_n,\bm{k})=\frac{1}{2}\sum_{a=\pm}\frac{\sigma_0+ a \bm{\sigma}\cdot\hat{\bm{k}}}{i\omega_n + \mu -a \varepsilon_{\bm{k}}}.
\ee
Using this expression and the phonon propagator $D^{0}_{\alpha\alpha'}(i\w_n,\bm{q})$ in \eqqref{eq:GreenPhonon}, we rewrite the expression in \eqqref{app:self} as
\ba
\Sigma(i\omega_m,\bm{k})&=\sum_{\alpha,a,b}\int_{|\bq|<k_D^{\alpha}}g^{\alpha}_{\bm{-q}}b\Lambda_{a}(\bm{k}-\bm{q})I^{\alpha}_{ab}(\bm{k},\bm{q},i\w_m)g^{\alpha}_{\bm{q}} \label{app:selfenergy3},
\ea
in terms of $\Lambda_{\pm}(\bm{k}) \equiv (\s^0 \pm  {\bm\s}\cdot\hat{\bm{k}})/2$ and 
\ba
I^{\alpha}_{ab}(\bm{k},\bm{q},i\w_m)&\equiv-\frac{1}{\beta}\sum_{i\w_n}\frac{1}{i\omega_{m}+\mu-i\w_n-a\varepsilon_{\bm{k}-\bm{q}}}\cdot\frac{1}{i\w_n-b\omega^{\alpha}_{\bm{q}}}\nonumber\\
&=\frac{1}{2\pi i}\oint_{C}\text{d} z \frac{N_{\text{B}}(z)}{(i\omega_{m}+\mu-z-a\varepsilon_{\bm{k}-\bm{q}})(z-b\omega^{\alpha}_{\bm{q}})},\label{app:sum}
\ea
where we converted the Matsubara sum into a contour integral in the complex plane over the Bose distribution function $N_{\text{B}}(z) \equiv (e^{\beta z} - 1)^{-1}$. To make the poles more explicit, we rewrite the integrand in \eqqref{app:sum} as
\ba
&\frac{N_{\text{B}}(z)}{(i\omega_{m}+\mu-z-a\varepsilon_{\bm{k}-\bm{q}})(z-b\omega^{\alpha}_{\bm{q}})} = \frac{N_{\text{B}}(z)}{i\omega_m+\mu-(a\varepsilon_{\bm{k}-\bm{q}}+b\omega^{\alpha}_{\bm{q}})}\left(\frac{1}{z-b\omega^{\alpha}_{\bm{q}}}-\frac{1}{z-i\omega_m-\mu+a\varepsilon_{\bm{k}-\bm{q}}}\right).
\ea
The integrand thus has poles at $z_1=b\omega^{\alpha}_{\bm{q}}$ and $z_2=i\omega_m+\mu-a\varepsilon_{\bm{k}-\bm{q}}$. Performing the integral in \eqqref{app:sum} using the residue theorem, we find
\be
I^{\alpha}_{ab}(\bm{k},\bm{q},i\w_m)=\frac{N_{\text{B}}(b\omega^{\alpha}_{\bm{q}})-N_{\text{B}}(i\omega_m +\mu-a\varepsilon_{\bm{k}-\bm{q}})}{i\omega_m+\mu-a\varepsilon_{\bm{k}-\bm{q}}-b\omega^{\alpha}_{\bm{q}}}. \label{eq:integralab}
\ee
As $i\w_m$ is a fermionic Matsubara frequency, the second term can be simplified by noting that 
\be
N_{\text{B}}(i\omega_{m}+\mu-a\varepsilon_{\bm{k}-\bm{q}})=-N_{\text{F}}(\mu-a\varepsilon_{\bm{k}-\bm{q}}),
\ee
where $N_{\text{F}}(x) \equiv (e^{\beta x} + 1)^{-1}$ is the Fermi-Dirac distribution function. Furthermore, we have the identities
\begin{subequations}
\ba
N_{\text{B}}(-\omega^{\alpha}_{\bm{q}})&=-1-N_{\text{B}}(\omega^{\alpha}_{\bm{q}}), \\
N_{\text{F}}(\mu-\varepsilon_{\bm{k}}) &=1-N_{\text{F}}(\varepsilon_{\bm{k}}-\mu),
\ea
\end{subequations}
such that the four terms defined by \eqqref{eq:integralab} read
\begin{subequations}
\ba
I^{\alpha}_{++}(\bm{k},\bm{q},i\w_m)&=\frac{N_{\text{B}}(\omega^{\alpha}_{\bm{q}})-N_{\text{F}}(\varepsilon_{\bm{k}-\bm{q}}-\mu)+1}{i\omega_m+\mu-\varepsilon_{\bm{k}-\bm{q}}-\omega^{\alpha}_{\bm{q}}}, \\
I^{\alpha}_{+-}(\bm{k},\bm{q},i\w_m)&=\frac{N_{\text{B}}(\omega^{\alpha}_{\bm{q}})+N_{\text{F}}(\varepsilon_{\bm{k}-\bm{q}}-\mu)}{i\omega_m+\mu-\varepsilon_{\bm{k}-\bm{q}}+\omega^{\alpha}_{\bm{q}}},\\
I^{\alpha}_{-+}(\bm{k},\bm{q},i\w_m)&=\frac{N_{\text{B}}(\omega^{\alpha}_{\bm{q}})+N_{\text{F}}(\varepsilon_{\bm{k}-\bm{q}}+\mu)}{i\omega_m+\mu+\varepsilon_{\bm{k}-\bm{q}}-\omega^{\alpha}_{\bm{q}}},\\
I^{\alpha}_{--}(\bm{k},\bm{q},i\w_m)&=\frac{N_{\text{B}}(\omega^{\alpha}_{\bm{q}})-N_{\text{F}}(\varepsilon_{\bm{k}-\bm{q}}+\mu)+1}{i\omega_m+\mu+\varepsilon_{\bm{k}-\bm{q}}+\omega^{\alpha}_{\bm{q}}}.
\ea
\end{subequations}
We now proceed by going to the zero-temperature limit, $T=0$, and we assume a positive chemical potential, $\mu>0$. In this limit there are no thermally activated phonons and fermions. Therefore $N_{\text{B}}(\omega^{\alpha}_{\bm{q}})=0$ and $N_{\text{F}}(\varepsilon_{\bm{k}-\bm{q}}+\mu)= 0$ such that $I^{\alpha}_{-+}(\bm{k},\bm{q},i\w_m) = 0$. Performing the analytical continuation to real frequencies $i\w_m \rightarrow \w + i\eta$, with $\eta\rightarrow0$, we obtain for the retarded self-energy
\ba
\Sigma(\omega,\bm{k})&=\sum_{\alpha}\int_{|\bq|<k_D^{\alpha}}g^{\alpha}_{\bm{-q}}\bigg[\frac{\Lambda_{+}(\bm{k}-\bm{q})\big[1-\vartheta(\mu - \varepsilon_{\bm{k}-\bm{q}})\big]}{\omega-\varepsilon_{\bm{k}-\bm{q}}-\omega^{\alpha}_{\bm{q}}+i\eta} +\frac{\Lambda_{+}(\bm{k}-\bm{q}) \vartheta(\mu - \varepsilon_{\bm{k}-\bm{q}})}{\omega-\varepsilon_{\bm{k}-\bm{q}}+\omega^{\alpha}_{\bm{q}}+i\eta}+\frac{\Lambda_{-}(\bm{k}-\bm{q})}{\omega+\varepsilon_{\bm{k}-\bm{q}}+\omega^{\alpha}_{\bm{q}}+i\eta}\bigg]g^{\alpha}_{\bm{q}},\label{app:selfenergy4}
\ea
which is exactly \eqqref{eq:Selfenergy2}. Note that we shifted $\w + \mu \rightarrow \w$ in \eqqref{app:selfenergy4}, such that all frequencies are measured with respect to the chemical potential. The obtained expression for the self-energy contains four contributions: two proportional to $\vartheta(\mu-\varepsilon_{\bm{k}-\bm{q}})$ and two with just a $1$ in the numerator. The former two are associated with the matter content and the latter two, being non-zero even at $\mu=0$, are vacuum contributions. 
\subsection{Sum over polarizations}
\noindent The next step is to perform the sum over the polarizations by using the vertex from \eqqref{eq:Vertex}. For briefness we introduce $\hat{p}^l \equiv (k-q)^l / |\bm{k}-\bm{q}|$ and $\tilde{g}^2_1 \equiv v_F^2 g_1^2 / 4\rho$. For the longitudinal contribution to the self-energy we then find
\ba
g^{L}_{-{\bm q}} \Lambda_{\pm}({\bm k}-{\bm q})g^{L}_{{\bm q}} &= \frac{\tilde{g}_1^2|{\bm q}|^2}{4\w_L({\bm q})}\lambda_a \lambda_b \hat{e}^L_{r}\hat{e}^L_s\big(\hat{q}_i\delta_{ar} + \hat{q}_a\delta_{ir}\big)\big(\hat{q}_j\delta_{bs} + \hat{q}_b \delta_{js}\big)\s^i\big(\s^0 \pm \s^l\hat{p}_l\big)\s^j \nonumber \\
&=\frac{\tilde{g}_1^2({\bm \lambda}\cdot{\bm q})^2}{\w_L({\bm q})}\bigg[\s^0 \pm \frac{\big[2{\bm k}\cdot\bm{q} - \bm{q}^2\big](\bm{q}\cdot\bm{\s})}{|{\bm k}-{\bm q}||\bm{q}|^2} \mp \frac{({\bm k}\cdot{\bm \s})}{|\bm{k}-\bm{q}|}\bigg], \label{eq:SumPolL}
\ea
where we used $\hat{e}^L_i = \hat{q}_i$ and the identity $\s^i\s^j = \delta^{ij} + i\varepsilon^{ijk}\s_k$ several times. The first term in the expression above is proportional to $\s^0$ and leads to $\Sigma_0^L(\w,\bm{k})$, whereas the other terms yield $\Sigma^L_{1,i}(\w,\bm{k})\s^i$. Doing the same for the sum over the transverse modes, we find
\ba
\sum_{\alpha = T_1,T_2}g^{\alpha}_{-\bm{q}} \Lambda_{\pm}(\bm{k}-\bm{q})g^{\alpha}_{\bm{q}} &= \frac{\tilde{g}_1^2|\bm{q}|^2}{4\w_T(\bm{q})}\lambda_a \lambda_b (\delta_{rs} - \hat{q}_r\hat{q}_s)\big(\hat{q}_i\delta_{ar} + \hat{q}_a\delta_{ir}\big)\big(\hat{q}_j\delta_{bs} + \hat{q}_b \delta_{js}\big)\s^i\big(\s^0 \pm \s^l\hat{p}_l\big)\s^j \nonumber \\
&=\frac{\tilde{g}_1^2}{4\w_T(\bm{q})}\big[\bl^2\bm{q}^2+(\bl\cdot\bm{q})^2\big]\s^0 \pm\frac{\tilde{g}_1^2}{4\w_T(\bm{q})|\bm{k}-\bm{q}|}\big[(\bl\cdot\bm{q})^2-\bl^2\bm{q}^2 \big](\bm{k}\cdot\bm{\s})  \nonumber\\
&\phantom{=}\pm\frac{\tilde{g}_1^2}{4\w_T(\bm{q})|\bm{k}-\bm{q}|}\bigg\{2\big[\bl^2- 4(\bl\cdot\hat{\bm{q}})^2\big](\bm{k}\cdot\bm{q})+2(\bl\cdot\bm{q})(\bl\cdot\bm{k})+5({\bm \lambda}\cdot\bm{q})^2-{\bm \lambda}^2\bm{q}^2\bigg\}(\bm{q}\cdot{\bm \s}) \nonumber\\
&\phantom{=}\pm\frac{\tilde{g}_1^2({\bm \lambda}\cdot\bm{q})\big[\bm{k}\cdot\bm{q}- \bm{q}^2\big]}{2\w_T(\bm{q})|\bm{k}-\bm{q}|}({\bm \lambda}\cdot{\bm \s}) \label{eq:SumPolT}.
\ea
\subsection{Tensorial decomposition}
\noindent We now proceed with the computation of the self-energy that has been made dimensionless by scaling all parameters on $|\bm{\lambda}|$, as we discussed in the paper. Additionally, we take for simplicity $\veps_L = \veps_T \equiv \veps$ and $k_D^L = k_D^T \equiv k_D$ in the following. Using \eqqref{eq:SumPolL} and \eqqref{app:selfenergy4}, we find for the longitudinal contributions to the dimensionless self-energy 
\begin{subequations}
\ba
\Sigma^L_0(\w,\bm{k}) &=  \hat{\lambda}^i\hat{\lambda}^j \int_{k_D>|\bm{q}|}\frac{q^iq^j}{|\bm{q}|}H_+(\w,\bm{k}-\bm{q},\bm{q}), \label{app:selfenergyL0} \\ 
\Sigma^L_{1,i}(\w,\bm{k})\s^i &= \hat{\lambda}^i\hat{\lambda}^j\s^l\int_{k_D>|\bm{q}|}\frac{\hat{q}^i\hat{q}^j\hat{q}^l\big(2\bm{k}\cdot\bm{q} - |\bm{q}|^2\big)}{|\bm{k}-\bm{q}|}H_-(\w,\bm{k}-\bm{q},\bm{q})  -(\bm{\s}\cdot\bm{k})\hat{\lambda}^i\hat{\lambda}^j\int_{k_D>|\bm{q}|}\frac{q^i\hat{q}^j}{|\bm{k}-\bm{q}|} H_-(\w,\bm{k}-\bm{q},\bm{q}) \label{app:selfenergyL1},
\ea
\end{subequations}
in terms of the function $H_{\pm}(\w, \bx,\by)$ defined in \eqqref{eq:HCollective}. Note that $\Sigma_0(\w,\bm{k})$ depends on $H_+(\w,\bm{x},\bm{y})$, whereas $\Sigma_{1,i}(\w,\bm{k})\s^i$ depends on $H_-(\w, \bm{x},\bm{y})$. This is ultimately due to the fact that the last term in \eqqref{app:selfenergy4} is proportional to $\Lambda_-(\bm{k}-\bm{q})$, whereas the other terms are proportional $\Lambda_+(\bm{k}-\bm{q})$.

In order to simplify the expression \eqqref{app:selfenergyL0}, we note that the integrand is symmetric in $i\leftrightarrow j$ and must therefore be a linear combination of $\delta^{ij}$ and $\hat{k^i}\hat{k^j}$. Making the ansatz $I^{ij}(k) = \alpha(k)\delta^{ij} +\beta(k) \hat{k}^i\hat{k}^j$ for the tensor resulting from the integral, one finds $\alpha(k) = \big(\delta_{ij} - \hat{k}_i\hat{k}_j\big)I^{ij}(k)/2$ and $\beta(k) = \big(3\hat{k}_i\hat{k}_j - \delta_{ij}\big)I^{ij}(k)/2$. After shifting the integration variable $\bm{q}\rightarrow\bm{q}+\bm{k}$, we find
\ba
&\Sigma^L_0(\w,\bm{k}) =  \frac{1}{2}\int_{k_D>|\bm{k}+\bm{q}|}\bigg\{\bm{q}^2 - (\hat{\bm{k}}\cdot\bm{q})^2 + (\hat{\bl} \cdot\hat{\bm{k}})^2\big[3(\hat{\bm{k}}\cdot\bm{q})^2 + 2\bm{k}^2 -\bm{q}^2 + 4\bm{k}\cdot\bm{q}\big]\bigg\}\frac{H_+(\w,\bm{q},\bm{k}+\bm{q}) }{|\bm{k}+\bm{q}|}. 
\ea
Only the terms on the second line depend on the angle between $\bm{\lambda}$ and $\bm{k}$. In the special cases $\bm{\lambda} \perp \bm{k}$ and $\bm{\lambda}\parallel \bm{k}$, we denote the self-energy as $\Sigma^{L,\perp}_0(\w,\bm{k})$ and $\Sigma^{L,\perp}_0(\w,\bm{k})$ and find 
\begin{subequations}
\ba
&\hspace{-.2cm}\Sigma^{L,\perp}_0(\w,\bm{k}) =  \frac{1}{2}\int_{k_D>|\bm{k}+\bm{q}|}\!\big[\bm{q}^2 - (\hat{\bm{k}}\cdot\bm{q})^2\big]\frac{H_+(\w,\bm{q},\bm{k}+\bm{q}) }{|\bm{k}+\bm{q}|}, \label{app:sigmaL0perpfin}\\
&\hspace{-.1cm}\Sigma^{L,\parallel}_0(\w,\bm{k}) =  \int_{k_D>|\bm{k}+\bm{q}|}\!\big[\bm{k}^2 + (\hat{\bm{k}}\cdot\bm{q})^2 + 2(\bm{k}\cdot\bm{q})\big]\frac{H_+(\w,\bm{q},\bm{k}+\bm{q}) }{|\bm{k}+\bm{q}|}. \label{app:sigmaL0parafin}
\ea
\end{subequations}
We follow a similar strategy for \eqqref{app:selfenergyL1}. The second integrand is of the same form as the integrand of \eqqref{app:selfenergyL0} and can thus be computed in the same way. The first integrand is symmetric in $i\leftrightarrow j\leftrightarrow l$. We can thus decompose the result of the corresponding integral as $I^{ijl}(k) = \alpha(k)\big(\delta^{ij}\hat{k}^l + \delta^{il}\hat{k}^j + \delta^{jl}\hat{k}^i\big) + \beta(k) \hat{k}^i\hat{k}^j\hat{k}^l$. The coefficients $\alpha(k)$ and $\beta(k)$ follow by contracting $I^{ijl}(k)$ with $\delta^{ij}\hat{k}^l$ and $\hat{k}^i\hat{k}^j\hat{k}^l$. They read $\alpha(k) = \big(\delta_{ij}\hat{k}_l - \hat{k}_i\hat{k}_j\hat{k}_l\big)I^{ijl}(k)/2$ and $\beta(k) = \big(5\hat{k}_i\hat{k}_j\hat{k}_l - 3\delta_{ij}\hat{k}_l\big)I^{ijl}(k)/2$. Combining all contributions we write $\Sigma_{1,i}^{L,\perp}(\w,\bm{k})\s^i = \Sigma_1^{L,\perp}(\w,\bm{k})(\hat{\bm{k}}\cdot\bm{\s})$ when $\bm{\lambda}\perp\bm{k}$ and find
\ba
\Sigma_1^{L,\perp}(\w,\bm{k}) &= \frac{1}{2}\int_{k_D > |\bm{k} + \bm{q}|}\big[(\hat{\bk}\cdot\hat{\bq})^2-1\big]\big[\big(\bm{q}^2 + \bm{k}^2\big)(\hat{\bm{k}}\cdot\hat{\bm{q}}) + 2|\bm{k}||\bm{q}|\big]\frac{\bm{q}^2H_-(\w,\bm{q},\bm{k}+\bm{q})}{|\bm{k}+\bm{q}|^3} , \label{app:sigmaL1perpfin}
\ea
In the case that $\bm{\lambda}\parallel\bm{k}$, we write $\Sigma_{1,i}^{L,\parallel}(\w,\bm{k})\s^i = \Sigma_1^{L,\parallel}(\w,\bm{k})(\hat{\bm{k}}\cdot\bm{\s})$, with
\ba
\Sigma_1^{L,\parallel}(\w,\bm{k}) &=-\int_{k_D > |\bm{k} + \bm{q}|}\!\bigg[ \bm{q}^2(\bm{q}^2 + \bm{k}^2)(\hat{\bm{k}}\cdot\hat{\bm{q}})^3 + 2|\bm{k}||\bm{q}|(2\bm{q}^2+ \bm{k}^2)(\hat{\bm{k}}\cdot\hat{\bm{q}})^2  \nonumber\\
&\phantom{=}\qquad\qquad+\bm{k}^2(\bm{k}^2 + 5\bm{q}^2)(\hat{\bm{k}}\cdot\hat{\bm{q}}) + 2\bm{k}^3|\bm{q}|\bigg]\frac{H_-(\w,\bm{q},\bm{k}+\bm{q})}{|\bm{k}+\bm{q}|^3}. \label{app:sigmaL1parafin}
\ea
Following the same procedure as for the longitudinal part, but starting from Eq.\eqref{eq:SumPolT} instead of Eq.\eqref{eq:SumPolL}, we find for transverse contribution
\begin{subequations}
\ba
\Sigma_0^{T,\perp}(\w,\bm{k}) &= \int_{k_D > |\bm{k} + \bm{q}|}\!\Big[3\bm{q}^2 + 4\bm{k}\cdot\bm{q} + 2\bm{k}^2 - (\hat{\bm{k}}\cdot\bm{q})^2 \Big]\frac{H_+(\w,\bm{q},\bm{k}+\bm{q})}{8|\bm{k}+\bm{q}|}, \label{app:sigmaT0perpfin} \\
\Sigma_0^{T,\parallel}(\w,\bm{k}) &=\int_{k_D > |\bm{k} + \bm{q}|}\!\Big[\bm{q}^2 + 4\bm{k}\cdot\bm{q} + 2\bm{k}^2 + (\hat{\bm{k}}\cdot\bm{q})^2 \Big]\frac{H_+(\w,\bm{q},\bm{k}+\bm{q})}{4|\bm{k}+\bm{q}|}\label{app:sigmaT0parafin} .
\ea
\end{subequations}
In the perpendicular case $\bm{\lambda}\perp\bm{k}$ we write similarly to the longitudinal case $\Sigma_{1,i}^{T,\perp}(\w,\bm{k})\s^i = \Sigma_1^{T,\perp}(\w,\bm{k})(\hat{\bm{k}}\cdot\bm{\s})$. Following the same steps that led to \eqqref{app:sigmaL1perpfin}, we obtain
\ba
\Sigma_1^{T,\perp}(\w,\bk) &= \int_{k_D > |\bm{k} + \bm{q}|}\!\bigg[2|\bm{k}|\bm{q}^3 -2|\bm{k}|\bm{q}^3(\hat{\bm{k}}\cdot\hat{\bm{q}})^4 - \bm{q}^2(5\bm{q}^2 + \bm{k}^2)(\hat{\bm{k}}\cdot\hat{\bm{q}})^3  -2|\bm{k}||\bm{q}|(\bm{k}^2 + 4\bm{q}^2)(\hat{\bm{k}}\cdot\hat{\bm{q}})^2 -2\bm{k}^4(\hat{\bm{k}}\cdot\hat{\bm{q}}) \nonumber\\
&\phantom{=}\qquad\qquad\qquad-\bm{q}^2(11\bm{k}^2 - 3\bm{q}^2)(\hat{\bm{k}}\cdot\hat{\bm{q}}) \bigg]\frac{H_-(\w,\bm{q},\bm{k}+\bm{q})}{8|\bm{k}+\bm{q}|^3}\label{app:sigmaT1perpfin}  ,
\ea
Finally, when $\bm{\lambda}\parallel\bm{k}$ we denote $\Sigma_{1,i}^{T,\parallel}(\w,\bm{k})\s^i = \Sigma_1^{T,\parallel}(\w,\bm{k})(\hat{\bm{k}}\cdot\bm{\s})$ and find
\ba
\Sigma_1^{T,\parallel}(\w,\bm{k}) &=\int_{k_D > |\bm{k} + \bm{q}|}\!\bigg[2|\bm{k}|\bm{q}^3(\hat{\bm{k}}\cdot\hat{\bm{q}})^4 + 5\bm{q}^2(\bm{k}^2 + \bm{q}^2)(\hat{\bm{k}}\cdot\hat{\bm{q}})^3  +2|\bm{k}||\bm{q}|(5\bm{q}^2 + 2\bm{k}^2)(\hat{\bm{k}}\cdot\hat{\bm{q}})^2 +  4|\bm{k}||\bm{q}|(\bm{k}^2 - \bm{q}^2)  \nonumber\\
&\phantom{=}\qquad\quad\qquad+(7\bm{k}^2\bm{q}^2 + 2\bm{k}^4 - 3\bm{q}^4)(\hat{\bm{k}}\cdot\hat{\bm{q}})\bigg]\frac{H_-(\w,\bm{q},\bm{k}+\bm{q})}{4|\bm{k}+\bm{q}|^3}.\label{app:sigmaT1parafin} 
\ea
We have now obtained all longitudinal and transverse contributions to the self-energy in the distinct cases $\bm{\lambda} \perp \bm{k}$ and $\bm{\lambda}\parallel \bm{k}$. Because we took $\veps_L = \veps_T \equiv \veps$ we can simplify the full expression for the self-energy by adding the longitudinal and transverse contributions and arrive at \eqqref{eq:Selfenergy4}. Using Eqs.\,\eqref{app:sigmaL0perpfin}, \eqref{app:sigmaL0parafin}, \eqref{app:sigmaT0perpfin} and \eqref{app:sigmaT0parafin} we obtain the functions $f_0^{\perp,\parallel}(\bm{k},\bm{q})$. They are given by
\begin{subequations}
\ba
f_0^{\perp}(\bm{k},\bm{q}) &= 7\bm{q}^2 -5(\hat{\bm{k}}\cdot\bm{q})^2 + 4(\bm{k}\cdot\bm{q}) + 2\bm{k}^2, \\
f_0^{\parallel}(\bm{k},\bm{q}) &= 2\bm{q}^2 + 10(\hat{\bm{k}}\cdot\bm{q})^2 + 24(\bm{k}\cdot\bm{q}) + 12\bm{k}^2.
\ea
\end{subequations}
Similarly, we obtain from Eqs.\,\eqref{app:sigmaL1perpfin},\eqref{app:sigmaL1parafin}, \eqref{app:sigmaT1perpfin} and \eqref{app:sigmaT1parafin} the functions $f_1^{\perp,\parallel}(\bm{k},\bm{q})$. They are given by
\begin{subequations}
\ba
f_1^{\perp}(\bm{k},\bm{q}) &= 3 k^2 q^2 x (x^2-5) -2 k^4 x-2 k^3 q (x^2+3)  -2 k q^3 (x^4+3)-q^4 x (x^2+1),  \\
f_1^{\parallel}(\bm{k},\bm{q}) &= 2k^2 q^2 x (x^2-13) -4 k^4 x -8 k^3 q (x^2+1)+ 4 k q^3 (x^4-3 x^2-2) + 2q^4 x(x^2-3),
\ea
\end{subequations}
where we introduced $x\equiv \hat{\bk}\cdot\hat{\bq}$.
\section{Additional figures}
\begin{figure}[htpb!]
\includegraphics[scale=.53]{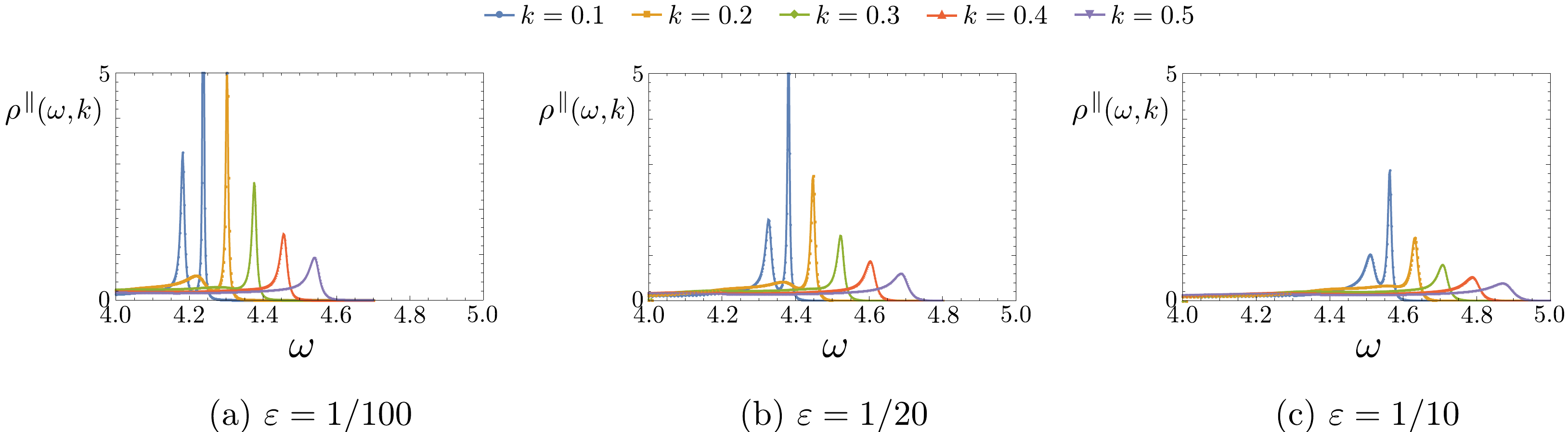}
\caption{Plots of the spectral function in the case $\bl \parallel\bk$, as a function of the frequency $\w$ and for five different values of $k = |\bk|/|\bl|$. In all three plots we used $k_D/\mu = 4.0$ and $G_1 = 1$.}
\label{fig:Eps}
\end{figure}

\noindent To supplement the discussion in the main text, we show in \figgref{fig:Eps} plots of the spectral function $\rho^{\parallel}(\w,\bk)$ for three different values for the ratio of the phonon and electron velocity $\veps = c /v_F$. From the figure it becomes clear that both the height and width, and the location of the satellite excitations is determined by $\veps$. The satellite peak is well-defined when $\veps\ll 1$ and $|\bk|/|\bl|$ is sufficiently small.

\end{document}